\documentclass[aps,prd,preprint,showpacs,showkeys]{revtex4}
\usepackage{graphics} 
\newcommand{\V}{\mathcal{V}}
\newcommand{\e}{\mathrm{e}}
\newcommand{\ep}{\epsilon}

\newcommand{\Sp}{\mathrm{Sp}~}
\def\slash#1{\hbox{$#1$\kern-0.5em\raise0.3ex\hbox{/}}}
\begin{document}
\preprint{KOBE-TH-03-05}  \title{Feynman graph solution to \\Wilson's
exact renormalization group} \author{Hidenori Sonoda}
\affiliation{Physics Department, Kobe University, Kobe 657-8501,
Japan} \email[E-mail address: ]{sonoda@phys.sci.kobe-u.ac.jp}
\begin{abstract}
We introduce a new prescription for renormalizing Feynman diagrams.
The prescription is similar to BPHZ, but it is mass independent, and
works in the massless limit as the MS scheme with dimensional
regularization.  The prescription gives a diagrammatic solution to
Wilson's exact renormalization group differential equation.
\end{abstract}
\maketitle

The purpose of this paper is to introduce a new renormalization
prescription for Feynman diagrams.  The prescription works for any
theory that permits a perturbative treatment in terms of Feynman
diagrams, irrespective of spacetime dimensions and statistics of
particles.  The prescription is for extracting the ultraviolet (UV)
finite part of any Feynman diagram.  It works for any theory, whether
renormalizable or not, but of course the prescription does not render
a non-renormalizable theory renormalizable.

Out of many possible renormalization prescriptions, two
renormalization schemes stand out.  One is the BPHZ scheme
\cite{BPHZ}, and the other is the minimal subtraction (MS) scheme with
dimensional regularization.\cite{thv,th73} The advantage of the former
(especially the improvement made by Zimmermann) is that necessary UV
subtractions are made at the level of integrands, and that Feynman
diagrams are automatically UV finite.  The advantage of the latter is
its ease with concrete calculations and its mass independence.

Our new prescription is justified by the exact renormalization group
of Wilson, and it shares some nice properties of the two schemes
mentioned above, in particular, the automatic UV finiteness and mass
independence.  The only drawback of the new method is that the
symmetry of the theory is not necessarily manifest.  Global linear
symmetry can be incorporated manifestly, but gauge symmetry and
nonlinearly realized symmetry must be enforced by hand.\footnote{It is
not as bad as it sounds.  Thanks to the exact renormalization group,
the study of Ward identities to all orders in perturbation theory is
relatively simple.\cite{s03b}}

In this paper we will mainly consider a real scalar field theory in
four dimensional euclidean space.  The propagator of a real scalar
particle with squared mass $m^2$ is given by $\frac{1}{p^2 + m^2}$.
We can decompose this into two parts:
\begin{equation}
\frac{1}{p^2 + m^2} = \frac{K(p \e^{-t})}{p^2 + m^2} + \frac{1 -
K(p \e^{-t})}{p^2 + m^2} \label{decomposition}
\end{equation}
where $K(p)$ is a non-negative smooth function of $p^2$ with the
property
\begin{equation}
K(p) = \cases{1 & for $p^2 < 1$\cr 0 & for $p^2 > 2^2$\cr}
\end{equation}
The first term of Eq.~(\ref{decomposition}) corresponds to the
propagation of low momentum fluctuations, and the second to that of
high momentum fluctuations.  We have chosen the scale of
renormalization as $\e^t$, where $t$ is an arbitrary logarithmic scale
parameter.\footnote{We could have used $\e^t \mu$ instead of $\e^t$.
But we don't.}  The choice of a particular form of $K(p)$ is not
important in the rest of the discussion except that it must be $1$ at
low momentum, and that it vanishes sufficiently fast at high momentum.

Our aim is to introduce a prescription for calculating Feynman
diagrams in which all the propagators are replaced by the high
momentum propagator $\frac{1-K(p \e^{-t})}{p^2+m^2}$.  Suppose
$n$-point vertex functions $\V_n (-t; p_1,\cdots,p_n)$ have been
defined as the sum of all possible Feynman diagrams with $n$ external
lines with momenta $p_1, \cdots, p_n$ for which all the internal
propagators are the high momentum propagators.  Then, the full Green
functions can be calculated using the low momentum propagator
$\frac{K(p \e^{-t})}{p^2+m^2}$ and the vertices $\V_n (-t)$.  In other
words the Green functions of the theory can be fully reproduced by the
``perfect action'' given by \footnote{$\int_p \equiv \int
\frac{d^4p}{(2\pi)^4}$}
\begin{eqnarray}
S [-t; \phi] &=& \int_p \frac{1}{2} \phi (p) \phi (-p) \frac{p^2 +
m^2}{K(p \e^{-t})}\nonumber\\ &&- \sum_{n=1}^\infty \int_{p_1 + \cdots
+ p_n = 0} \frac{1}{n!} \phi (p_1) \cdots \phi (p_n) \, \V_n (-t; p_1,
\cdots, p_n)
\end{eqnarray}
The high momentum fluctuations have already been incorporated into the
vertices, and they do not propagate anymore.  Despite the lack of
explicit high momentum fluctuations, the perfect action describes the
physics of the continuous space.\footnote{Strictly speaking, the
perfect action reproduces the full Green functions only for external
momenta less than $1$.  We can remove this restriction, however, by
introducing an ad hoc rule that we use the standard propagator
$\frac{1}{p^2+m^2}$ instead of $\frac{K(p)}{p^2 + m^2}$ for the most
external lines.}

There are two types of Feynman diagrams: one-particle reducible (1PR)
graphs and one-particle irreducible (1PI) graphs.  Given a 1PR graph,
we can define it as the product of two subgraphs and a high momentum
propagator as in FIG.~1.
\begin{figure}
\includegraphics{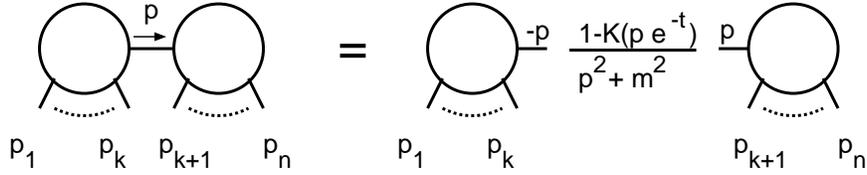}
\caption{A 1PR graph is reduced to a product of two subgraphs.  The
momentum conservation implies $p \equiv p_1 + \cdots + p_k = -(p_{k+1}
+ \cdots + p_n)$.}
\end{figure}
We can keep reducing a 1PR graph to a multiple product of 1PI graphs
and high momentum propagators with fixed momenta.  Hence, our task is
reduced to defining 1PI graphs.

We need a renormalization prescription to define 1PI graphs in which
all loop momenta are larger than $\e^t$.  We adopt an incremental
procedure: given a 1PI graph, we define its value by integrating over
the loop momenta scale by scale from $\e^t$ all the way to infinity.

As a preparation for this incremental procedure, let us first make the
following observation.  The high momentum propagator can be decomposed
further as follows:
\begin{equation}
\frac{1 - K(p \e^{-t})}{p^2 + m^2} = \int_t^\infty dt' \,\frac{\Delta (p
\e^{-t'})}{p^2 + m^2} \label{slices}
\end{equation}
where we define
\begin{equation}
\Delta (p \e^{-t}) \equiv \frac{\partial}{\partial t} K (p \e^{-t})
\end{equation}
We must use $K(0) = 1$ to derive Eq.~(\ref{slices}).  The physical
meaning of Eq.~(\ref{slices}) is clear if we recall that $\Delta (p
\e^{-t})$ is nonvanishing only for $p$ of order $\e^{t}$ (FIG.~2): the
integrand on the right hand side of Eq.~(\ref{slices}) gives the
contribution from the momentum of order $\e^{t'}$ per unit logarithmic
scale $t'$.
\begin{figure}
\includegraphics{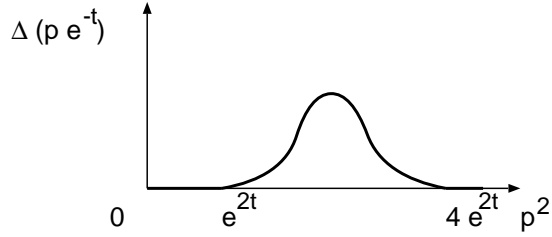}
\caption{$\Delta (p \e^{-t})$ is non-vanishing only for $p^2$ of order
$\e^{2t}$.}
\end{figure}

Given a 1PI diagram $G$, any of its internal lines belongs to a loop,
and its momentum is integrated over.  By cutting an internal line, we
generate a diagram $G'$, not necessarily 1PI, with one less number of
loops but with the same number of elementary interaction vertices.  In
order to define the 1PI diagram recursively, we assume that the
diagrams of lower order, i.e., those with less number of either loops
or elementary vertices, have been defined already.  We can then define
the 1PI graph $G$ by
\begin{eqnarray}
&&\V_G (-t; p_1, \cdots, p_n) \nonumber\\ &\equiv& \sum_{G'} \Bigg[\,
\int_t^\infty dt' \int_q \, \Delta (q \e^{-t'}) \Bigg\lbrace \,
\frac{1}{q^2+m^2}\, \V_{G'} (-t'; q, -q, p_1, \cdots, p_n)\nonumber\\
&&\qquad\qquad\qquad\qquad\qquad\quad - \Gamma_{y_{G'}+2} \left(
\frac{1}{q^2 + m^2} \, \V_{G'} (-t'; q, - q, p_1, \cdots, p_n) \right)
\Bigg\rbrace \nonumber\\ && - \int^t dt' \,\e^{2t'}\, \int_q \Delta
(q) \cdot \Gamma_{y_{G'}+2} \left( \frac{1}{q^2 + m^2 \e^{-2t'}} \,
\V_{G'} (-t'; q \e^{t'}, - q \e^{t'}, p_1, \cdots, p_n) \right)
\,\Bigg] \label{G}
\end{eqnarray}
where $G'$ is a graph obtained by cutting one internal line of $G$,
and we must sum over all possible distinct choices of an internal
line.  We must multiply a symmetry factor of $\frac{1}{2}$ if $G'$
cannot be distinguished from the corresponding graph with $q$ and $-q$
interchanged.  This is necessary in order to avoid overcounting of the
phase space.  The definitions of the other unexplained symbols will be
given shortly.
\begin{figure}
\includegraphics{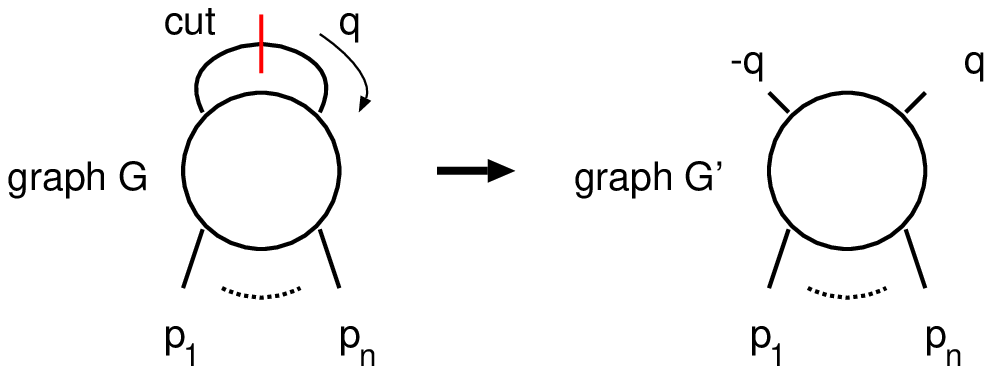}
\caption{Any internal line of a 1PI graph belongs to a loop.}
\end{figure}

First about the scale dimension $y_{G'}$.  This is defined by the
asymptotic behavior of the vertex $\V_{G'} (-t)$ as $t \to \infty$:
\begin{equation}
\V_{G'} (-t; q \e^t, -q \e^t, p_1, \cdots, p_n) \longrightarrow
\text{O} \left( \e^{y_{G'} t} \right)
\end{equation}
where we take two momenta (loop momenta for $G$) as order $\e^t$.  In
a renormalizable theory, such as $\phi^4$, the scale dimension is
determined by the number of external legs ($n+2$ for $G'$) so that
\begin{equation}
y_{G'} = 4 - (n + 2)
\end{equation}
But $y_{G'}$ could be larger than this for non-renormalizable
theories.  Since the internal momenta of $\V_{G'} (-t)$ are all larger
than $\e^t$, we can expand $\V_{G'} (-t)$ in powers of the mass $m$
and external momenta $p_i~(i=1,\cdots,n)$ with no infrared
problem\footnote{The absence of the IR divergence is the main
advantage of the new prescription over the BPHZ scheme.}:
\begin{eqnarray}
&&\V_{G'} (-t; q \e^t, -q \e^t, p_1, \cdots, p_n)\nonumber\\ &=&
\e^{y_{G'} t}~A_{G'} (-t; q) + \e^{(y_{G'}-2)t}~m^2 B_{G'} (-t; q) +
\e^{(y_{G'}-2)t}\, \frac{1}{2} \sum_{i,j} p_i \cdot p_j\, C_{G'; ij}
(- t; q) + \cdots
\end{eqnarray}
where we have taken the angular average over $q$.  The coefficients of
the Taylor expansion, $A_{G'}(-t; q)$, $B_{G'}(-t;q)$, $C_{G'}(-t;q)$,
etc., are all of order $1$, or to be more precise, finite degree
polynomials of $t$, to be explained later.  It is crucial to observe
that each power of $m$ or $p_i$ costs a power of $\e^{-t}$.  This is
because any loop momentum for $\V_{G'} (-t)$ is at least of order
$\e^t$, and the expansion is in powers of the ratio of $m$ or $p_i$ to
$\e^t$.

We define the symbol $\Gamma_y$ by the finite sum of the above Taylor
series up to (and including) the $y$-th order terms.  For example,
\begin{eqnarray}
&&\Gamma_2 \,\V_{G'} (-t; q \e^t, - q \e^t, p_1, \cdots, p_n)
\nonumber\\ &\equiv& \e^{y_{G'} t} A_{G'} (-t; q) + \e^{(y_{G'}-2)t}
m^2 B_{G'} (-t; q) + \e^{(y_{G'}-2)t} \frac{1}{2} \sum_{i,j} p_i \cdot
p_j C_{G'; ij} (- t; q)
\end{eqnarray}
Similarly, with $q$ rescaled by $\e^{-t}$, we obtain
\begin{eqnarray}
&&\Gamma_2 \left( \frac{1}{q^2 + m^2} \V_{G'} (-t; q, - q, p_1,
\cdots, p_n) \right) \nonumber\\ &\equiv& \e^{y_{G'}t} \frac{1}{q^2}
A_{G'} (-t; q \e^{-t}) + \e^{(y_{G'}-2)t} m^2 \left( \frac{1}{q^2}
B_{G'} (-t; q \e^{-t}) - \frac{\e^{2t}}{q^4} A_{G'} (-t; q
\e^{-t})\right) \nonumber\\ && + \e^{(y_{G'}-2)t} \frac{1}{2}
\sum_{i,j} p_i \cdot p_j \frac{1}{q^2} C_{G'; ij} (- t; q \e^{-t})
\end{eqnarray}

Now that $\Gamma_{y_{G'}+2}$ has been defined, let us look at the
first two integrals of the definition (\ref{G}).  Rescaling the loop
momentum $q$, the first loop integral is given by
\begin{eqnarray}
I_{G'} (-t') &\equiv& \int_q \Delta (q \e^{-t'}) 
\frac{1}{q^2+m^2} \, \V_{G'} (-t'; q,-q,p_1,\cdots,p_n)\nonumber\\
&=& \e^{2 t'}
\int_q \Delta (q) \, \frac{1}{q^2 + m^2 \e^{-2t'}} \,
\V_{G'} (-t'; q \e^{t'}, - q \e^{t'}, p_1, \cdots, p_n)  \label{I}
\end{eqnarray}
In the second expression, the range of the momentum integral is
restricted to $q$ of order $1$, and the integral is finite, free from
UV or IR divergences.  Similarly, the second loop integral is given by
\begin{eqnarray}
J_{G'} (-t') &\equiv& \int_q \Delta (q \e^{-t'}) \Gamma_{y_{G'}+2}
\left( \frac{1}{q^2+m^2}\, \V_{G'} (-t'; q,-q,p_1,\cdots,p_n) \right)
\nonumber\\
&=& \e^{2 t'} \int_q \Delta (q)
\Gamma_{y_{G'}+2} \left( \frac{1}{q^2+m^2 \e^{-2t'}}\, \V_{G'} (-t'; q
\e^{t'},-q \e^{t'},p_1,\cdots,p_n) \right)\nonumber\\
&=& \Gamma_{y_{G'}+2} I_{G'} (-t') \label{J}
\end{eqnarray}
The Taylor expansion commutes with integration over $q$, and $J_{G'}
(-t')$ gives the finite sum of the Taylor series of $I_{G'} (-t')$ up
to order $y_{G'} + 2$.

As $t' \to \infty$, we obtain the asymptotic behavior
\begin{eqnarray}
I_{G'} (-t') &=& \e^{(y_{G'}+2) t'} \int_q \frac{\Delta (q)}{q^2}
A_{G'} (-t'; q) + \e^{y_{G'} t'} m^2 \int_q \frac{\Delta (q)}{q^2}
\left( B_{G'} (-t'; q) - \frac{A_{G'} (-t';q)}{q^2} \right)\nonumber\\
&& + \e^{y_{G'} t'} \frac{1}{2} \sum_{i,j} p_i \cdot p_j \int_q
\frac{\Delta (q)}{q^2} C_{G';ij} (-t'; q) + \cdots
\end{eqnarray}
Each power of $m$ or $p_i$ costs a power of $\e^{-t'}$.  Hence, the
integral $J_{G'} (-t')$ corresponds only to the part of $I_{G'} (-t')$
that does not vanish in the limit $t' \to \infty$.  Therefore, we
obtain
\begin{equation}
I_{G'} (-t') - J_{G'} (-t') = \mathrm{O} \left( \e^{-t'} {t'}^k \right)
\end{equation}
where $k$ is an integer.\footnote{$k$ is at most the number of loops
in $G'$.}  This implies that we can integrate the difference $I_{G'}
(-t') - J_{G'} (-t')$ over $t'$ all the way to $\infty$.  Therefore,
the integral
\begin{equation}
\int_t^\infty dt'\, \left( I_{G'} (-t') - J_{G'} (-t') \right)
\end{equation}
is free of UV divergences, and well defined.

The finite counterterms, given by the last integral in Eq.~(\ref{G}),
are introduced to cancel the $t$ dependence of the UV subtraction
$J_{G'} (-t)$.  Let us call the finite counterterms by $F_{G'} (-t)$.
We want $F_{G'} (-t)$ to satisfy
\begin{equation}
\frac{\partial}{\partial t} F_{G'} (-t) = - J_{G'} (-t) \label{tdep}
\end{equation}
so that 
\begin{equation}
- \frac{\partial}{\partial t} \left[ \int_t^\infty dt'\, \left( I_{G'}
(-t') - J_{G'} (-t') \right) + F_{G'} (-t) \right] = I_{G'} (-t) 
\end{equation}
Eq.~(\ref{tdep}) is formally solved by
\begin{equation}
F_{G'} (-t) = - \int^t dt'\, J_{G'} (-t') \label{F}
\end{equation}
We recall that the terms of $J_{G'}(-t')$ have the form $\e^{y t'}
t'^k$ where $y \ge 0$ and $k$ is a non-negative integer. To make the
definition (\ref{F}) precise, we need to specify what we mean by the
finite integral of $\e^{y t'} t'^k$.

There is no unique choice for the finite integrals.  Any specification
is a convention free of any physical meaning.  We find it most
convenient to introduce our version of a minimal subtraction scheme by
using the following convention: for $y > 0$
\begin{equation}
\int^t dt'\, \e^{y t'} {t'}^k \equiv \e^{y t} P_{y,k} (t) \label{finiteone}
\end{equation}
where $P_{y,k}$ is the $k$-th degree polynomial of $t$ defined
uniquely by
\begin{equation}
\frac{d}{dt} \left( \e^{y t} P_{y,k} (t) \right) = \e^{y t} t^k \label{P}
\end{equation}
and for $y=0$
\begin{equation}
\int^t dt' \,{t'}^k \equiv \int_0^t dt'\,{t'}^k = \frac{t^{k+1}}{k+1}
\label{finitetwo}
\end{equation}
With this convention, Eq.~(\ref{F}) defines the finite counterterms.

To summarize, and using a more symbolic notation than Eq.~(\ref{G}),
we define
\begin{equation}
\V_G (-t; p_1, \cdots, p_n) = \sum_{G'} \left[ \int_t^{\infty} dt'\,
(I_{G'} (-t') - J_{G'} (-t')) + F_{G'} (-t) \right] \label{symbolic}
\end{equation}
where the summation is over all distinct choices of an internal line
of $G$, $I_{G'}$ is given by Eq.~(\ref{I}), $J_{G'}$ by Eq.~(\ref{J}),
and $F_{G'}$ by Eq.~(\ref{F}) and the convention (\ref{finiteone},
\ref{P}, \ref{finitetwo}).

The prescription (\ref{G}) or (\ref{symbolic}) is reminiscent of the
BPHZ scheme in which Taylor expansions in powers of external momenta
are used.  There is a significant difference between the two schemes,
however. In BPHZ the whole vertex functions are expanded, but here
only the high energy part of the vertex functions are expanded.  This
is why our prescription works for the massless theory, but not BPHZ.

A remark is in order.  We have mentioned that the Taylor coefficients
of $\V_{G'}(-t)$, i.e., $A_{G'}(-t), B_{G'}(-t), C_{G'}(-t)$, etc.,
are finite polynomials of $t$.  We have also mentioned that $J_{G'}
(-t)$ consists of terms of the form $\e^{y t} t^k$.  These two
statements are the same.  To prove either statement, we notice that
the definition (\ref{symbolic}) implies the asymptotic behavior
\begin{equation}
\V_G (-t) \longrightarrow \sum_{G'} F_{G'} (-t)
\end{equation}
since
\begin{equation}
\int_t^\infty dt' \left( I_{G'} (-t') - J_{G'} (-t') \right)
\longrightarrow 0
\end{equation}
as $t \to \infty$.  Now, $F_{G'} (-t)$ is a finite integral (over $t$)
of $J_{G'} (-t)$, which is determined by the asymptotic behavior of
$\V_{G'} (-t)$.  Therefore, the asymptotic behavior of $\V_G (-t)$ is
determined by that of $\V_{G'} (-t)$.  Hence, it is not hard to see
that we can prove the above mentioned $t$-dependence of $J_{G'} (-t)$
by mathematical induction on the order (number of loops plus
elementary vertices) of graphs.

As an example, we take the four-dimensional $\phi^4$ theory.  (See
FIG.~4.)  
\begin{figure}
\includegraphics{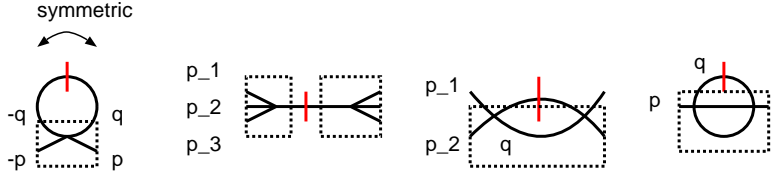}
\caption{Examples from $\phi^4$ theory}
\end{figure}
Let $(- \lambda)$ be the elementary vertex.  Then, the one-loop self
energy is given by
\begin{eqnarray}
\V_2^{(1)} (-t) &=& (- \lambda) \Bigg[ \, \int_t^\infty dt'
\frac{1}{2} \int_q \Delta (q \e^{-t'}) \left (\frac{1}{q^2 + m^2} -
\frac{1}{q^2} + \frac{m^2}{q^4} \right)\nonumber\\ && \quad -
\frac{\e^{2t}}{4} \int_q \frac{\Delta (q)}{q^2} + \frac{m^2}{2} t
\int_q \frac{\Delta (q)}{q^4} \, \Bigg]
\end{eqnarray}
which is independent of external momentum.  The symmetry factor
$\frac{1}{2}$ is necessary, since the cut graph with external momenta
$q, -q$ is the same as that with $-q, q$.

At order $\lambda^2$, the six-point vertex is given by 1PR graphs, and
we obtain
\begin{equation}
\V_6 (-t; p_1, \cdots, p_6) = (- \lambda)^2 \left[ \frac{1 - K((p_1 +
p_2 + p_3) \e^{-t})}{(p_1 + p_2 + p_3)^2 + m^2} +
\text{9~permutations} \right]
\end{equation}
The one-loop four-point vertex in the s-channel is given by
\begin{eqnarray}
&&\V_{4\mathrm{s}}^{(2)} (-t; p_1, \cdots, p_4)\nonumber\\ &=&
(-\lambda)^2 \Bigg[ \int_t^\infty dt' \int_q \, \Delta (q \e^{-t'})
\Bigg\{ \frac{1}{q^2 + m^2} \, \frac{1 - K((q + p_1 +
p_2)\e^{-t'})}{(q+p_1 + p_2)^2 + m^2} - \frac{1 - K(q \e^{-t'})}{q^4}
\Bigg\} \nonumber\\ &&\qquad\quad- t \int_q \frac{\Delta (q) (1 -
K(q))}{q^4} \: \Bigg] \label{fourpoint}
\end{eqnarray}

To compute the two-point vertex $\V_2$ at order $\lambda^2$, we need
to expand $\V_{4\mathrm{s}}^{(2)}$ in external momentum $p$ and
squared mass:
\begin{equation}
\Gamma_{2} \V_{4\mathrm{s}}^{(2)} (-t; q\e^t ,p, - q\e^t, - p) = A_4
(-t; q) + \e^{-2t} \left( m^2 B_4 (-t; q) + p^2 C_4 (-t; q) \right)
\label{expansion}
\end{equation}
where $A_4 (-t;q), B_4 (-t; q), C_4 (-t; q)$ are at most linear in $t$
for fixed $q$.\footnote{We have taken the average over the direction
of $q$, and ignored the terms proportional to $q\cdot p$.}  Then we
obtain
\begin{eqnarray}
&&\V_2^{(2)} (-t; p,-p) \nonumber\\ &=& (-\lambda)^2 \Bigg[
\int_t^\infty dt' \,\e^{2t'} \int_q \Delta (q) \Bigg\lbrace
\frac{1}{q^2 + m^2 \e^{-2t'}} \V_{4\mathrm{s}}^{(2)} (-t'; q\e^{t'},
p, -q\e^{t'}, - p)\nonumber\\ && - \frac{1}{q^2} \left( 1 -
\frac{\e^{-2t'} m^2}{q^2} \right) A_4 (-t'; q) - \frac{\e^{-2t'}
m^2}{q^2} B_4 (-t'; q) - \frac{\e^{-2t'} p^2}{q^2} C_4 (-t'; q)
\Bigg\rbrace\nonumber\\ && - \int^t dt'\, \e^{2t'} \int_q \frac{\Delta
(q)}{q^2} A_4 (-t'; q) - m^2 \int_0^t dt'\, \int_q \,{\Delta (q)}
\left( \frac{B_4 (-t';q)}{q^2} - \frac{A_4 (-t'; q)}{q^4} \right)
\nonumber\\ &&\qquad - p^2 \int_0^t dt'\, \int_q \frac{\Delta
(q)}{q^2} C_4 (-t'; q) \, \Bigg]
\end{eqnarray}
We will give more explicit expressions of $A_4, B_4, C_4$ in Appendix
A.

So far we have only described the prescription.  A justification is in
order.  The recursive definition (\ref{G}) is constructed so that for
a 1PI graph $G$, we find
\begin{equation}
- \frac{\partial}{\partial t} \V_{G} (-t; p_1,\cdots, p_n) = \int_q
\frac{\Delta (q \e^{-t})}{q^2+m^2} \, \sum_{G'} \V_{G'} (-t; q, - q,
p_1, \cdots, p_n) \label{dVdt}
\end{equation}
Recall Eq.~(\ref{tdep}): the finite subtraction $F_{G'} (-t)$ has been
introduced to cancel the $t$-dependence of the UV subtraction $J_{G'}
(-t)$.

Now, summing over all Feynman diagrams with $n$ external lines,
including both 1PR and 1PI diagrams, we obtain the $n$-point vertex:
\begin{equation}
\V_n (-t; p_1, \cdots, p_n) \equiv
\sum_{G~\mathrm{with~}n~\mathrm{legs}} \V_{G} (-t; p_1, \cdots, p_n)
\end{equation}
Eq.~(\ref{dVdt}) and the rule for 1PR graphs imply
\begin{eqnarray}
&&- \frac{\partial}{\partial t} \V_n (-t; p_1, \cdots, p_n)
  \nonumber\\ &=& \sum_{k=0}^{\left[\frac{n}{2}\right]}
  \sum_{\mathrm{partitions}~\sigma} \V_{k+1} (-t; p_{\sigma (1)},
  \cdots, p_{\sigma (k)}, - (p_{\sigma (1)} + \cdots + p_{\sigma
  (k)})) \nonumber\\ && \quad \times \frac{\Delta (p_{\sigma (1)} +
  \cdots + p_{\sigma (k)})}{(p_{\sigma (1)} + \cdots + p_{\sigma
  (k)})^2 + m^2}\, \V_{n-k+1} (-t; p_{\sigma (k+1)}, \cdots, p_{\sigma
  (n)}, p_{\sigma (1)} + \cdots + p_{\sigma (k)}) \nonumber\\ && +
  \frac{1}{2} \int_q \frac{\Delta (q \e^{-t})}{q^2+m^2} \, \V_{n+2}
  (-t; q, - q, p_1, \cdots, p_n) \label{ERG}
\end{eqnarray}
where the sum is over all possible partitions of $n$ external momenta
into two groups.  This is the exact renormalization group (ERG)
differential equation of Wilson \cite{wk74} in the form given by
Polchinski.\cite{pol84} The exact renormalization group equation
guarantees that the Green functions are independent of the choice of
the scale parameter $t$.  The prescription (\ref{G}) can then be
understood as the diagrammatic solution of the exact RG differential
equation (\ref{ERG}).  Hence, the exact renormalization group
justifies our diagrammatic prescription (\ref{G}).

Actually our prescription is more than a diagrammatic solution to the
ERG differential equation.  A new formulation of the ERG in terms of
integral equations has been derived recently in Ref.~\cite{s02}.  Our
renormalization prescription is a solution of the integral equations
in terms of Feynman diagrams.\footnote{In fact we have come up with
the new prescription by seeking for a diagrammatic solution of the
integral equations.}

A generalization to fermions is straightforward.  All we need to do is
to replace the propagator by
\begin{equation}
\frac{1}{\slash{p} + i m} \longrightarrow \frac{K(p)}{\slash{p} + i m}
\end{equation}
A classical test of a renormalization scheme is the derivation of the
axial anomaly.\footnote{The axial anomaly has already been computed
using a similar method of the exact renormalization group in
Refs.~\cite{bonini94, bonini98}.  Our regulator
$$
1 - (1-K(q\e^{-t}))(1-K((q+k_1)\e^{-t}))(1-K((q+k_1+k_2)\e^{-t}))
$$
for $T_{2,\mu\alpha\beta}$, given by Eq.~(\ref{Ttwo}), is replaced by
$K(q\e^{-t})K((q+k_1)\e^{-t})K((q+k_1+k_2)\e^{-t})$ in
Refs.~\cite{bonini94, bonini98}.}  (See FIG.~5.)  For the massless
fermion, we obtain the following amplitude\footnote{We take the
electric charge as $1$.  The overall minus sign is due to the Fermi
statistics.  We define $\gamma_5 \equiv \gamma_1 \gamma_2 \gamma_3
\gamma_4$ so that $\Sp \gamma_5 \gamma_\alpha \gamma_\beta
\gamma_\gamma \gamma_\delta = 4 \ep_{\alpha\beta\gamma\delta}$ where
$\ep_{1234} = 1$.}:
\begin{eqnarray}
&& \mathcal{T}_{\mu \alpha \beta} (k_1, k_2) = - \int_t^{\infty} dt'
\int_q\, \Sp \gamma_5 \gamma_\mu \frac{1}{\slash{q}} \gamma_\alpha
\frac{1}{\slash{q} + \slash{k}_1} \gamma_\beta
\frac{1}{\slash{q}+\slash{k}_1+\slash{k}_2} \nonumber\\ &&\times
\Bigg[\, \Delta (q \e^{-t'})
\{1-K((q+k_1)\e^{-t'})\}\{1-K((q+k_1+k_2)\e^{-t'})\} \nonumber\\ &&
\quad+ \{1 - K(q\e^{-t'})\} \Delta ((q+k_1)\e^{-t'})
\{1-K((q+k_1+k_2)\e^{-t'})\}\nonumber\\ && \quad + \{1 -
K(q\e^{-t'})\}\{1-K((q+k_1)\e^{-t'})\} \Delta ((q+k_1+k_2)\e^{-t'})\,
\Bigg] \nonumber\\ &-& \int_q \Sp \gamma_5 \gamma_\mu
\frac{1}{\slash{q}} \gamma_\alpha \frac{1}{\slash{q} + \slash{k}_1}
\gamma_\beta \frac{1}{\slash{q}+\slash{k}_1+\slash{k}_2}\nonumber\\ &&
\times \left[ 1 - \{1 - K(q\e^{-t})\} \{ 1 - K((q+k_1)\e^{-t}) \} \{ 1
- K((q+k_1+k_2)\e^{-t}) \} \right] \nonumber\\ &-& (\mbox{the above
two integrals with } k_1,\alpha \leftrightarrow k_2,\beta) + c\,
\ep_{\mu\alpha\beta\gamma} (k_1 - k_2)_\gamma
\end{eqnarray}
where $c$ is a constant coefficient of the finite counterterm.  This
is independent of the logarithmic scale parameter $t$.
\begin{figure}
\includegraphics{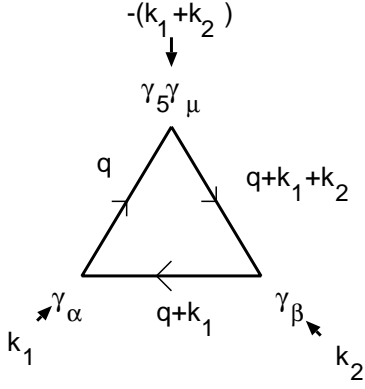}
\caption{Triangle anomaly}
\end{figure}
Potentially the integrand of the $t'$ integral behaves as $\e^{t'}$
for large $t'$, but we can check its absence.  In fact the integrand
behaves as $\e^{-t'}$, and no UV subtraction is necessary.  The
$t$-dependence of the first integral cancels that of the second, and
the whole right-hand side is independent of $t$.  We note that the
loop momentum $q$ can be shifted; a potential UV divergence come from
the integral over $t'$, not $q$.

In Appendix B we will show the following two:
\begin{enumerate}
\item For the current conservation
\begin{equation}
(k_1)_\alpha \mathcal{T}_{\mu\alpha\beta} (k_1,k_2) = 0
\end{equation}
we must choose
\begin{equation}
c = - \frac{8}{3 (4\pi)^2}
\end{equation}
\item The axial anomaly is given by
\begin{equation}
(k_1+k_2)_\mu \mathcal{T}_{\mu\alpha\beta}(k_1,k_2) =
\frac{8}{(4\pi)^2} \ep_{\alpha\beta\gamma\delta} (k_1)_\gamma
(k_2)_\delta
\end{equation}
\end{enumerate}
Hence, our prescription passes the classical test.

In conclusion we have given a new renormalization prescription of
Feynman diagrams.  The prescription gives a refinement of the naive
momentum cutoff regularization, and it resembles both BPHZ and MS
with dimensional regularization.  The prescription gives manifestly UV
finite expressions like the BPHZ scheme, and it is mass independent
and works for massless theories like the MS scheme with dimensional
regularization.  The exact renormalization group of Wilson validates
our renormalization prescription.

Gauge symmetry and non-linearly realized symmetry are not manifest
under our new renormalization scheme, and we must introduce additional
finite counterterms to enforce a symmetry.  With the help of the exact
renormalization group, however, the analysis of the relevant Ward
identities to all orders in perturbation theory becomes
straightforward.\cite{s03b}

\appendix

\section{More details on the two-loop self-energy}

Similar calculations can be found in Ref.~\cite{s03a}.  We omit the
factor of $(-\lambda)^2$ in the rest of the calculations.  From
Eqs.~(\ref{fourpoint}) we obtain
\begin{eqnarray}
&&\V_{4\mathrm{s}}^{(2)} (-t; q \e^t, p, - q \e^t, -p)\nonumber\\ &=&
\int_0^\infty dt' \int_{q'} \Delta (q' \e^{-t'}) \left\lbrace
\frac{1}{q'^2 + m^2 \e^{-2t}} \cdot \frac{1 - K((q' + q + p \e^{-t})
\e^{-t'})}{(q'+q+p \e^{-t})^2 + m^2 \e^{-2t}} - \frac{1 - K(q'
\e^{-t'})}{q'^4} \right\rbrace \nonumber\\ &&\quad - t \int_{q'}
\frac{\Delta (q')(1 - K(q'))}{q'^4}
\end{eqnarray}
Expanding this in powers of $m^2$ and $p^2$, we obtain
\begin{eqnarray}
&& A_4 (-t; q) = \int_0^\infty dt' \int_{q'} \frac{\Delta
(q'\e^{-t'})}{q'^2} \left( \frac{1 - K((q'+q)\e^{-t'})}{(q'+q )^2} -
\frac{1 - K(q' \e^{-t'})}{q'^2} \right)\nonumber\\ && \qquad\qquad - t
\int_{q'} \frac{\Delta (q') (1 - K(q'))}{{q'}^4}\\ &&B_4 (-t;
q)\nonumber\\ &=& - \int_0^\infty dt' \int_{q'} \Delta (q'
\e^{-t'}) (1 - K((q'+q)\e^{-t'})) \left( \frac{1}{{q'}^4 (q'
+ q)^2} + \frac{1}{q'^2 (q'+q)^4} \right)\\ &&C_4 (-t; q) = \int_0^\infty
dt' \frac{\partial}{\partial p^2} \left[ \int_{q'} \frac{\Delta (q'
\e^{-t'}) (1 - K((q'+q +p)\e^{-t'}))}{{q'}^2} \right]_{p^2=0}
\end{eqnarray}
This shows manifestly that $A_4 (-t; q)$ is at most linear in $t$, and
that both $B_4 (-t; q)$ and $C_4 (-t; q)$ are independent of $t$.  We
will write them as $B_4 (q)$ and $C_4 (q)$, respectively.

We can simplify the expression for $B_4$ further by integrating over
$t'$ first:
\begin{eqnarray}
B_4 (q) &=& \int_0^\infty dt' \frac{\partial}{\partial t'} \int_{q'}
\frac{(1-K(q' \e^{-t'}))(1 - K((q'+q)\e^{-t'}))} {{q'}^4 (q' + q)^2}
\nonumber\\ &=& - \int_{q'} \frac{(1-K(q'))(1 - K((q'+q)))} {{q'}^4
(q' + q)^2}
\end{eqnarray}
This integral is UV (and IR) finite.

Hence, using the above results we can compute the finite counterterms:
\begin{eqnarray}
&\textrm{(1)}&\int^t dt'\, \e^{2t'} \int_q \frac{\Delta (q)}{q^2} A_4
(-t'; q)\nonumber\\ &=& \frac{\e^{2t}}{2} \int_q \frac{\Delta
(q)}{q^2} \int_0^\infty dt' \int_{q'} \frac{\Delta (q'
\e^{-t'})}{{q'}^2} \left( \frac{1 - K((q'+q)e^{-t'})}{(q'+q)^2} -
\frac{1 - K(q' \e^{-t'})}{{q'}^2} \right) \nonumber\\ &&\qquad\qquad -
\e^{2t} \left( \frac{t}{2} - \frac{1}{4} \right) \int_q \frac{\Delta
(q)}{q^2} \int_{q'} \frac{\Delta (q') (1 - K(q'))}{{q'}^4}, \\
&\textrm{(2)}&\int_0^t dt' \, \int_q \Delta (q) \left( \frac{B_4
(q)}{q^2} - \frac{A_4 (-t'; q)}{q^4}\right) \nonumber\\ &=& - t \int_q
\frac{\Delta (q)}{q^2} \int_{q'} \frac{(1-K(q))(1 - K(q'+q))}{{q'}^4
(q'+q)^2} \nonumber\\ && - t \int_q \frac{\Delta (q)}{q^4}
\int_0^\infty dt' \int_{q'} \frac{\Delta (q' \e^{-t'})}{{q'}^2} \left(
\frac{1 - K((q'+q)e^{-t'})}{(q'+q)^2} - \frac{1 - K(q'
\e^{-t'})}{{q'}^2} \right)\nonumber\\ && + \frac{t^2}{2} \int_q
\frac{\Delta (q)}{q^4} \int_{q'} \frac{\Delta (q') (1 -
K(q'))}{{q'}^4}, \\ &\textrm{(3)}&\int_0^t dt'\, \int_q \frac{\Delta
(q)}{q^2} C_4 (q)\nonumber\\ &=& t\int_0^\infty dt' \int_q
\frac{\Delta (q)}{q^2} \frac{\partial}{\partial p^2} \left\lbrace
\int_{q'} \frac{\Delta (q' \e^{-t'})}{{q'}^2} \frac{1 -
K((q'+q+p)\e^{-t'})}{(q'+q+p)^2}\right\rbrace_{p=0}\nonumber\\ &=& t
\int_0^\infty dt' \int_q \frac{\Delta (q \e^{t'})}{q^2} \int_{q'}
\frac{\Delta (q')}{{q'}^2} \frac{d^2 K(q'+q)}{(d
(q'+q)^2)^2}\nonumber\\ &=& t \int_q \frac{K(q)}{q^2} \int_{q'}
\frac{\Delta (q')}{{q'}^2} \frac{d^2 K(q'+q)}{(d (q'+q)^2)^2}
\end{eqnarray} 
Some of the integrals have values independent of the choice of $K$,
but we will not discuss it here.

\section{Derivation of the axial anomaly}

We give details of the calculation of the axial anomaly in this
appendix.  We first define
\begin{eqnarray}
&& T_{1, \mu \alpha \beta} (-t; k_1, k_2) \equiv - \int_t^{\infty} dt'
\int_q\, \Sp \gamma_5 \gamma_\mu \frac{1}{\slash{q}} \gamma_\alpha
\frac{1}{\slash{q} + \slash{k}_1} \gamma_\beta \frac{1}{\slash{q}+
\slash{k}_1+\slash{k}_2}\nonumber\\ && \times \Bigg[\, \Delta (q
\e^{-t'}) \{1-K((q+k_1)\e^{-t'})\}\{1-K((q+k_1+k_2)\e^{-t'})\}
\nonumber\\ && \quad+ \{1 - K(q\e^{-t'})\} \Delta ((q+k_1)\e^{-t'})
\{1-K((q+k_1+k_2)\e^{-t'})\}\nonumber\\ && \quad + \{1 -
K(q\e^{-t'})\}\{1-K((q+k_1)\e^{-t'})\} \Delta ((q+k_1+k_2)\e^{-t'})\,
\Bigg]
\end{eqnarray}
and 
\begin{eqnarray}
&&T_{2, \mu\alpha\beta}(-t; k_1, k_2) \equiv - \int_q \, \Sp \gamma_5
\gamma_\mu \frac{1}{\slash{q}} \gamma_\alpha \frac{1}{\slash{q} +
\slash{k}_1} \gamma_\beta \frac{1}{\slash{q}+
\slash{k}_1+\slash{k}_2} \nonumber\\ && \times \left[ 1 - \{1 -
K(q\e^{-t})\} \{ 1 - K((q+k_1)\e^{-t}) \} \{ 1 - K((q+k_1+k_2)\e^{-t})
\} \right] \label{Ttwo}
\end{eqnarray}
so that 
\begin{eqnarray}
\mathcal{T}_{\mu\alpha\beta}(k_1,k_2) &=& T_{1,\mu\alpha\beta} (-t; k_1,k_2) +
T_{2,\mu\alpha\beta} (-t; k_1,k_2)\nonumber\\ && +
T_{1,\mu\beta\alpha} (-t; k_2, k_1) + T_{2,\mu\beta\alpha}
(-t; k_2,k_1) + c\, \ep_{\mu\alpha\beta\gamma} (k_1 - k_2)_\gamma
\end{eqnarray}
The $t$-dependence of $T_1$ cancels that of $T_2$, and $\mathcal{T}$ is
independent of $t$.

To compute $(k_1+k_2)_\mu T_{1,\mu\alpha\beta} (k_1,k_2)$, we use the
well-known trick:
\begin{equation}
\slash{k}_1 + \slash{k}_2 = \slash{q} - (\slash{q}-\slash{k}_1 -
\slash{k}_2)
\end{equation}
We also use the identity
\begin{eqnarray}
&&\Delta (q \e^{-t'})
\{1-K((q+k_1)\e^{-t'})\}\{1-K((q+k_1+k_2)\e^{-t'})\} \nonumber\\ &&
\quad+ \{1 - K(q\e^{-t'})\} \Delta ((q+k_1)\e^{-t'})
\{1-K((q+k_1+k_2)\e^{-t'})\}\nonumber\\ && \quad + \{1 -
K(q\e^{-t'})\}\{1-K((q+k_1)\e^{-t'})\} \Delta
((q+k_1+k_2)\e^{-t'})\nonumber\\ &=& \frac{\partial}{\partial t'}
\left[ 1 - \{1 - K(q \e^{-t'})\}
\{1-K((q+k_1)\e^{-t'})\}\{1-K((q+k_1+k_2)\e^{-t'})\}\right]
\end{eqnarray}
We then obtain
\begin{eqnarray}
&&(k_1+k_2)_\mu T_{1,\mu\alpha\beta} (-t; k_1, k_2)\nonumber\\ &=&
- \int_t^\infty dt' \frac{\partial}{\partial t'} \int_q \left( \Sp
\gamma_5 \gamma_\alpha \frac{1}{\slash{q}-\slash{k}_1} \gamma_\beta
\frac{1}{\slash{q}-\slash{k}_1-\slash{k}_2} + \Sp \gamma_5
\frac{1}{\slash{q}} \gamma_\alpha \frac{1}{\slash{q}-\slash{k}_1}
\gamma_\beta \right)\nonumber\\ &&\times 
\left[ 1 - \{1 - K(q \e^{-t'})\}
\{1-K((q+k_1)\e^{-t'})\}\{1-K((q+k_1+k_2)\e^{-t'})\} \right]
\end{eqnarray}
For the first trace, we replace $q$ by $-q+k_1+k_2$.  Then, we get
\begin{eqnarray}
&&(k_1+k_2)_\mu T_{1,\mu\alpha\beta} (-t; k_1, k_2)\nonumber\\ &=& -
\int_t^\infty \frac{\partial}{\partial t'} \int_q \Bigg[ \left\{ 1 -
(1 - K(q \e^{-t'})) (1-K((q+k_1)\e^{-t'}))(1-K((q+k_1+k_2)\e^{-t'}))
\right\}\nonumber\\ &&\qquad\qquad \times \Sp \gamma_5
\frac{1}{\slash{q}} \gamma_\alpha \frac{1}{\slash{q}-\slash{k}_1}
\gamma_\beta \quad - (k_1 \leftrightarrow k_2) \, \Bigg]
\end{eqnarray}
Hence,
\begin{eqnarray}
&&(k_1+k_2)_\mu T_{1,\mu\alpha\beta} (-t; k_1, k_2)\nonumber\\ &=& -
\lim_{T \to \infty} \int_q \Bigg[ \e^{2T} \left\{ 1 - (1-K(q))(1-K(q -
k_1 \e^{-T}))(1-K(q-(k_1+k_2)\e^{-T})) \right\} \nonumber\\ && \times
\Sp \gamma_5 \frac{1}{\slash{q}} \gamma_\alpha
\frac{1}{\slash{q}-\slash{k}_1 \e^{-T}} \gamma_\beta \quad - (k_1
\leftrightarrow k_2) \Bigg] \quad - (k_1+k_2)_\mu T_{2,\mu\alpha\beta}
(-t; k_1,k_2)
\end{eqnarray}
Therefore, we obtain
\begin{eqnarray}
&&(k_1+k_2)_\mu \left( T_{1, \mu\alpha\beta} (-t; k_1, k_2) + T_{2,
\mu\alpha\beta} (-t; k_1, k_2) \right) \nonumber\\ &=& - \lim_{T \to
\infty} \int_q \Bigg[ \e^{2T} \left\{ 1 - (1-K(q))(1-K(q - k_1
\e^{-T}))(1-K(q-(k_1+k_2)\e^{-T})) \right\} \nonumber\\ && \times \Sp
\gamma_5 \frac{1}{\slash{q}} \gamma_\alpha
\frac{1}{\slash{q}-\slash{k}_1 \e^{-T}} \gamma_\beta \quad - (k_1
\leftrightarrow k_2) \Bigg]
\end{eqnarray}
This expression shows clearly that the anomaly comes from the UV
limit.  Computing the trace
\begin{equation}
\Sp \gamma_5 \frac{1}{\slash{q}} \gamma_\alpha
\frac{1}{\slash{q}-\slash{k}_1 \e^{-T}} \gamma_\beta \nonumber\\
= - \frac{4 \e^{-T}}{q^2 (q-k_1 \e^{-T})^2} \ep_{\mu\alpha\nu\beta}
q_\mu k_{1 \nu}
\end{equation}
we obtain
\begin{eqnarray}
&&(k_1+k_2)_\mu \left(T_{1, \mu\alpha\beta} (-t; k_1,k_2) +
T_{2,\mu\alpha\beta} (-t; k_1,k_2) \right) \nonumber\\ &=&
\lim_{T\to\infty} \int_q \Bigg[ \e^{2T} \left\lbrace 1 - (1-K(q))^2 (1
- K(q - k_1 \e^{-t})(1 - K(q - k_2 \e^{-T}))\right\rbrace\nonumber\\
&&\qquad\qquad \times \frac{4 \e^{-T}}{q^2 (q-k_1 \e^{-T})^2}
\ep_{\mu\alpha\nu\beta} q_\mu k_{1,\nu} - (k_1 \leftrightarrow k_2)
\Bigg]
\end{eqnarray}
Note the first integral vanishes if we take $q_2 = 0$.  Hence,
expanding in powers of $q_1, q_2$, only the coefficient of $q_2$ gives
a nonvanishing result:
\begin{eqnarray}
&&(k_1+k_2)_\mu T_{\mu\alpha\beta} (k_1,k_2)\nonumber\\ &=& \int_q
\frac{(1-K(q))^2 \Delta (q)}{q^4} \left[ 4 \ep_{\mu\alpha\nu\beta}
\frac{(q \cdot k_2) q_\mu k_{1\nu}}{q^2} - (k_1 \leftrightarrow k_2)
\right]\nonumber\\ &=& 2 \ep_{\alpha\beta\mu\nu} k_{1\mu} k_{2\nu}
\int_q \frac{(1-K(q))^2 \Delta (q)}{q^4}
\end{eqnarray}
Since
\begin{equation}
\Delta (q) = - 2 q^2 \frac{d K(q)}{d q^2}
\end{equation}
we obtain
\begin{equation}
\int_q \frac{(1-K(q))^2 \Delta (q)}{q^4}
= \frac{1}{24 \pi^2} \int_0^\infty dq^2 \frac{d}{dq^2} (1-K(q))^3 =
\frac{1}{24 \pi^2}
\end{equation}

Finally, we obtain
\begin{equation}
(k_1+k_2)_\mu \left( T_{1, \mu\alpha\beta} (-t; k_1, k_2) + T_{2,
\mu\alpha\beta} (-t; k_1, k_2) \right) = \frac{4}{3}
\frac{1}{(4\pi)^2} \ep_{\alpha\beta\mu\nu} k_{1\mu} k_{2\nu}
\end{equation}
A similar calculation gives
\begin{equation}
k_{1\alpha} \left( T_{1, \mu\alpha\beta} (-t; k_1, k_2) + T_{2,
\mu\alpha\beta} (-t; k_1, k_2) \right) = \frac{4}{3}
\frac{1}{(4\pi)^2} \ep_{\mu\beta\nu\tau} k_{1\nu} k_{2\tau}
\end{equation}

To recapitulate, we have obtained
\begin{eqnarray}
(k_1+k_2)_\mu \mathcal{T}_{\mu\alpha\beta} (k_1,k_2) &=& \left(
\frac{8}{3} \frac{1}{(4\pi)^2} - 2 c \right) \ep_{\alpha\beta\mu\nu}
k_{1\mu} k_{2\nu}\\
k_{1\alpha} \mathcal{T}_{\mu\alpha\beta} (k_1,k_2) &=& \left(
\frac{8}{3} \frac{1}{(4\pi)^2} + c \right) \ep_{\mu\beta\nu\tau}
k_{1\nu} k_{2\tau}
\end{eqnarray}
For the latter to vanish (conservation of the vector current), we must
choose
\begin{equation}
c = - \frac{8}{3} \frac{1}{(4\pi)^2}
\end{equation}
and we obtain the axial anomaly:
\begin{equation}
(k_1+k_2)_\mu \mathcal{T}_{\mu\alpha\beta} (k_1,k_2) =
\frac{8}{(4\pi)^2} \ep_{\alpha\beta\mu\nu} k_{1\mu} k_{2\nu}
\end{equation}

\begin{acknowledgments}
This work was partially supported by the Grant-In-Aid for Scientific
Research from the Ministry of Education, Culture, Sports, Science, and
Technology, Japan (\#14340077).
\end{acknowledgments}

\bibliography{feynman}

\begin{thebibliography}{10}
\expandafter\ifx\csname natexlab\endcsname\relax\def\natexlab#1{#1}\fi
\expandafter\ifx\csname bibnamefont\endcsname\relax
  \def\bibnamefont#1{#1}\fi
\expandafter\ifx\csname bibfnamefont\endcsname\relax
  \def\bibfnamefont#1{#1}\fi
\expandafter\ifx\csname citenamefont\endcsname\relax
  \def\citenamefont#1{#1}\fi
\expandafter\ifx\csname url\endcsname\relax
  \def\url#1{\texttt{#1}}\fi
\expandafter\ifx\csname urlprefix\endcsname\relax\def\urlprefix{URL }\fi
\providecommand{\bibinfo}[2]{#2}
\providecommand{\eprint}[2][]{\url{#2}}

\bibitem[{\citenamefont{Zimmermann}(1970)}]{BPHZ}
\bibinfo{author}{\bibfnamefont{W.}~\bibnamefont{Zimmermann}}, in
  \emph{\bibinfo{booktitle}{Lectures on Elementary Particles and Quantum Field
  Theory}} (\bibinfo{publisher}{MIT Press}, \bibinfo{year}{1970}),
  vol.~\bibinfo{volume}{1}, pp. \bibinfo{pages}{395--589}.

\bibitem[{\citenamefont{'t~Hooft and Veltman}(1972)}]{thv}
\bibinfo{author}{\bibfnamefont{G.}~\bibnamefont{'t~Hooft}} \bibnamefont{and}
  \bibinfo{author}{\bibfnamefont{M.}~\bibnamefont{Veltman}},
  \bibinfo{journal}{Nucl.~Phys.} \textbf{\bibinfo{volume}{B44}},
  \bibinfo{pages}{189} (\bibinfo{year}{1972}).

\bibitem[{\citenamefont{'t~Hooft}(1973)}]{th73}
\bibinfo{author}{\bibfnamefont{G.}~\bibnamefont{'t~Hooft}},
  \bibinfo{journal}{Nucl.~Phys.} \textbf{\bibinfo{volume}{B61}},
  \bibinfo{pages}{455} (\bibinfo{year}{1973}).

\bibitem[{\citenamefont{Wilson and Kogut}(1974)}]{wk74}
\bibinfo{author}{\bibfnamefont{K.~G.} \bibnamefont{Wilson}} \bibnamefont{and}
  \bibinfo{author}{\bibfnamefont{J.}~\bibnamefont{Kogut}},
  \bibinfo{journal}{Phys.~Repts.} \textbf{\bibinfo{volume}{12C}},
  \bibinfo{pages}{75} (\bibinfo{year}{1974}).

\bibitem[{\citenamefont{Polchinski}(1984)}]{pol84}
\bibinfo{author}{\bibfnamefont{J.}~\bibnamefont{Polchinski}},
  \bibinfo{journal}{Nucl.~Phys.} \textbf{\bibinfo{volume}{B231}},
  \bibinfo{pages}{269} (\bibinfo{year}{1984}).

\bibitem[{\citenamefont{Sonoda}(2003)}]{s02}
\bibinfo{author}{\bibfnamefont{H.}~\bibnamefont{Sonoda}},
  \bibinfo{journal}{Phys.~Rev.} \textbf{\bibinfo{volume}{D67}},
  \bibinfo{pages}{065011} (\bibinfo{year}{2003}).

\bibitem[{\citenamefont{M.~Bonini}(1994)}]{bonini94}
\bibinfo{author}{\bibfnamefont{G.~M.} \bibnamefont{M.~Bonini},
  \bibfnamefont{M.~D'Attanasio}}, \bibinfo{journal}{Phys.~Lett.}
  \textbf{\bibinfo{volume}{B329}}, \bibinfo{pages}{249} (\bibinfo{year}{1994}).

\bibitem[{\citenamefont{M.~Bonini}(1998)}]{bonini98}
\bibinfo{author}{\bibfnamefont{F.~V.} \bibnamefont{M.~Bonini}},
  \bibinfo{journal}{Nucl.~Phys.} \textbf{\bibinfo{volume}{B511}},
  \bibinfo{pages}{479} (\bibinfo{year}{1998}).

\bibitem[{\citenamefont{Sonoda}({\natexlab{a}})}]{s03b}
\bibinfo{author}{\bibfnamefont{H.}~\bibnamefont{Sonoda}}, \eprint{work in
  progress}.

\bibitem[{\citenamefont{Sonoda}({\natexlab{b}})}]{s03a}
\bibinfo{author}{\bibfnamefont{H.}~\bibnamefont{Sonoda}},
  \eprint{hep-th/0302044 (to be revised with a more complete list of
  references)}.

\end{thebibliography}

\end{document}